\documentclass[a4paper,10pt,aps]{revtex4-1}

\usepackage{amsmath}
\usepackage{bbm}
\usepackage{graphicx,amssymb,bm,latexsym,color,epsf}
\usepackage{breqn}
\usepackage[colorlinks]{hyperref} 

\newcommand{\be}{\begin{equation}}
\newcommand{\ee}{\end{equation}}
\newcommand{\bea}{\begin{eqnarray}}
\newcommand{\eea}{\end{eqnarray}}

\makeatletter
\let\cat@comma@active\@empty
\makeatother
\begin{document}

\title{On the regularization of Lifshitz-type field theories}

\author{Alfio Bonanno}
\email{alfio.bonanno@inaf.it}
\affiliation{INAF, Osservatorio Astrofisico di Catania, via S.Sofia 78, 
I-95123 Catania, Italy, \, and\\ 
INFN, Sezione di Catania, via S. Sofia 64, I-95123, Catania, Italy}

\author{Miok Park}
\email{ miokpark@kias.re.kr }
\affiliation{School of Physics, Korea Institute for Advanced Study, 
Seoul 02455, Republic of Korea}

\author{Les\l{}aw Rachwa\l{} }
\email{grzerach@gmail.com}
\affiliation{Departamento de F\'{\i}sica, ICE, Universidade Federal 
de Juiz de Fora,\\
Juiz de Fora,  36036-900, MG,  Brazil}

\author{Dario Zappal\`a}
\email{dario.zappala@ct.infn.it}
\affiliation{INFN, Sezione di Catania, via S. Sofia 64, I-95123, 
Catania, Italy}

\begin{abstract}
\vskip 30pt
\centerline{ABSTRACT}
\vskip 10pt
We consider Lifshitz-type scalar theories with explicit breaking of the 
Lorentz symmetry that, in addition, exhibit anisotropic scaling laws 
near the ultraviolet fixed point. 
Using the proper time regularization method on the spatial coordinates only,
we derive the regularized form of the one-loop effective potential in such
theories. We study the main features of the one-loop effective potential 
and, also, the RG flow of the scale-dependent potential both in 
the IR and UV regimes.
The beta functions for the couplings are derived.
\end{abstract} 
\maketitle

\setcounter{page}{2}

\section{Introduction}
Phase transitions associated to Lifshitz points, with their peculiar 
anisotropic scaling, were introduced and studied long ago, for the 
first time in \cite{Horn} and essentially applied 
to condensed matter problems \cite{hornrev,selke1988,Diehl}.
More recently the presence of Lifshitz points was investigated in
the high-energy realm, such as the electromagnetic field theory
\cite{horava:ym}, or the ultraviolet (UV) behavior of scalar 
fields \cite{Iengo,Kikuchi,eune}, or 
the more renowned Ho\v{r}ava-Lifshitz formulation  of the gravitational
theory \cite{horava}, subsequently generalized to black hole 
physics (see e.g. \cite{Cai,Eune:2010})
and cosmology (e.g. \cite{Brand,Son}).
 
The central issue related to the anisotropic Lifshitz points is the 
non-uniform scaling of the time and space coordinates, that can be 
summarized as
\begin{equation}
\label{eq:scaling}
t\to b^{z}\,t, \;\;\;\;\;\;\;\; x^{i}\to b\,x^{i},
\end{equation}
where $b$ is the rescaling parameter and $z$ is the critical scaling
exponent. This leads to the following non-uniform scaling dimensions 
$[t]_{s}=-z$ and $[x^{i}]_{s}=-1$ and to the peculiar derivative
sector, for instance in the action of a generic scalar field:
\begin{equation}
S=\!\int\!d^{3}x\,dt\left(\frac{1}{2}\dot{\phi}^{2}-\frac{1}{2}\alpha^{2}
\left(\partial_{i}^{z}\phi\right)^{2}-V\right),\label{eq: actionorig}
\end{equation}
where the dot indicates derivative with respect to the time variable, 
the index $i$ refers to the spatial coordinates,
$V(\phi)$ is the potential depending on the field $\phi$ only, and 
$\alpha$ is a dimensionful constant. 
Moreover, by the symbol $\partial^z_i$ we shall understand 
the $\frac{z}{2}$-th power of the spatially covariant 
Laplacian operator $\partial_i^2$.
The different scaling dimension of 
space and time coordinates requires a different number of  
derivatives with respect to these variables,
while the correct dimension is 
guaranteed by the constant parameter $\alpha$. 

The relative weight between the two derivative terms 
in (\ref{eq: actionorig}) 
(which is regulated by the parameter $\alpha$)
can generate a Lifshitz point that rules the UV physics of the model.
In particular, the value of the index $z$, being related to the power
of the spatial momentum in the propagator, is crucial to establish the
degree of divergence of the various diagrams and therefore the UV
structure of the model. On one side, a larger value of $z$ does soften 
the UV behavior, on the other side it produces a larger violation of the
Lorentz symmetry which is instead fully realized when $z=1$. 
So, for instance, 
the Ho\v{r}ava-Lifshitz formulation of gravity requires $z=3$, 
\cite{horava},
and one expects that the Renormalization Group (RG) flow towards 
the infrared  (IR) region  would modify $z$, so that $z\to 1$ and in 
the IR one eventually recovers a Lorentz invariant effective theory.

Incidentally, there is another way to treat all variables on the same
footing in order to recover a Lorentz invariant action, that is
to require the same non-standard scaling dimension for both time and
space variables, which is known as isotropic Lifshitz scaling
\cite{Horn,Diehl3}. 
 Thus, for instance, the simplest
isotropic Lifshitz  scaling is realized for a (Lorentz 
invariant) action containing terms quadratic in the field with 
four derivatives both in space and time. Isotropic Lifshitz points
show interesting properties according to the number of dimensions
\cite{Bonanno:2014yia,Zappala:2017vjf,Zappala:2018khg,Zapp,Defenu}, 
yet maintaining the same number of space and time derivatives of the field.

Actually, in general the explicit violation of Lorentz symmetry 
has an immediate
drawback related to the regularization scheme, which, in general,
is constructed accordingly to this symmetry, in such a way to
exploit its properties, thus resulting in a simple and elegant
procedure. If $z\neq 1$, Lorentz symmetry is lost and 
one has to choose a suitable regularization scheme to deal with the 
divergent integrals that appear both in the perturbative computations
and also in the development of the RG flow equations within this
approach. For instance, in \cite{eune}, the effective potential of 
Eq. (\ref{eq: actionorig}) with $z=2$ is computed by making the most 
straightforward choice for the  UV regulator, i.e. a sharp cutoff
on the spatial 3-momentum variable.

However, this sharp cutoff is known not to be suitable in 
the case of gauge and  gravity theories, as it conflicts with the
symmetries of the corresponding actions. 
Therefore, it is convenient to resort to a  more flexible 
regularization method that could preserve such symmetries.
To this purpose, we shall consider the proper time regulator 
\cite{Schwinger:1951nm}
that has been widely used both in the computation of the effective 
potential and of its RG improvements in the case of standard scaling 
\cite{Oles,Liao:1994fp,Liao:1995nm,Bohr,boza2001,Litpaw3,lit02,bola04},
and that has been recently revisited in \cite{dealwis18,Bonanno:2019ukb}
and, also, turns out to be more appropriate to treat gauge theories
\cite{Liao:1995nm}.
This kind of regulator  was used as well to evaluate the RG flow 
of a scalar theory in the case of isotropic Lifshitz scaling
\cite{Bonanno:2014yia},  and  we expect that it can
be suitably adapted to the anisotropic case. 
 In fact, an approach 
to the computation of the effective potential
similar to the one we shall discuss in this paper is presented
in \cite{Farias:2011aa},
although in our opinion not thoroughly elaborated.

In what follows, we will consider the toy model 
analyzed in \cite{eune}, 
i.e. the action in (\ref{eq: actionorig}) with $z=2$. 
In the action, in addition to the full potential $V$ 
that includes all higher order terms in the field $\phi$ 
(which, in this framework, has scaling dimension $[\phi]_{s}=1/2\,$), 
we retain the marginally deformed kinetic term only and 
neglect other renormalizable derivative operators that
would pointlessly complicate our analysis.
Since the theory naturally splits space and time, we shall 
deal with them separately.

In Section \ref{sec2}, we construct a proper time representation specifically for this 
case and  compute the one-loop effective potential.
In Section \ref{sec3}, we analyze some details of the renormalized one-loop effective
potential, while in  Section \ref{sec4} we determine the  RG flow that gives access
to the  $\beta$-functions of the various couplings. 
Our conclusions are reported in Section \ref{sec5}.

\section{Regularization scheme}
\label{sec2}

As a first step we compute the one-loop effective potential of
the action (\ref{eq: actionorig}) with the tree potential
\begin{equation}
V=\frac{m^{2}}{2}\phi^{2}+\sum_{n=1}^{4}\frac{\lambda_{n}}{(2n+2)!}\phi^{2n+2}\,,
\label{eq: potentialtree}
\end{equation}
that contains only the relevant (and marginal) powers 
of the field $\phi$, according to the non-standard scaling dimensions
outlined above. 

The one-loop computation involves integrals over the
four momentum components, but the different scaling of the 
space and time variables implies the breaking of the full Lorentz symmetry.
Therefore, in the loop integrals it is convenient to first perform 
the integral over $p^{0}$, which resembles the same integral 
in the standard case, and only later the integral 
over the spatial momenta $p^{i}$.

In our case of modified kinetic term, to the one-loop accuracy
and up to a field independent infinite constant,
the quantum effective potential in Minkowski spacetime 
takes the following form  
\begin{equation}
V_{{\rm 1l}}=-i \hbar \frac{1}{2}\!\int\!\frac{d^{4}p}{(2\pi)^{4}}\ln\frac{\delta^{2}S}{\delta\phi^{2}}
=-i \hbar \frac{1}{2}\!\int\!\frac{d^{4}p}{(2\pi)^{4}}\ln\left[p_{0}^{2}
-\alpha^{2}\left(\vec{p}^{\,2}\right)^{2}-V'{}' + i \epsilon\right]\,,
\label{eq: potentialdef}
\end{equation}
where $V'{}'$ is the second derivative of the potential with respect
to the field $\phi$ and $\epsilon$ is a positive constant which should be sent to zero in the end. After performing the $p^{0}$ integral 
we are left with the spatial integral

\begin{equation}
 V_{{\rm 1l}}=\frac{1}{2}\!\int\!\frac{d^{3}\vec{p}}{(2\pi)^{3}}
\sqrt{\alpha^{2}\left(\vec{p}^{\,2}\right)^{2}+V'{}'}\,,
\label{eq:potential3d}
\end{equation}
which, formally, is the same result of the standard approach, 
provided one
defines the energy $E$ in this case through the modified dispersion
relation 
$E^{2}=\alpha^{2}\left(\vec{p}^{\,2}\right)^{2}+V'{}'\,.$
Then, on a vanishing background the second
derivative of the potential $V'{}'$ is a constant and the UV 
divergences in our problem appear in the resolution of the integral
in (\ref{eq:potential3d}). The latter shows a $O(3)$ symmetry that 
can be used to integrate over the angular variables, to obtain
\begin{equation}
V_{{\rm 1l}}=\frac{1}{4\pi^{2}}\!\int\!dp\, 
p^{2}\sqrt{\alpha^{2}p^{4}+V'{}'}\,.\label{eq: integral}
\end{equation}

At this point, instead of treating  the UV divergences
by cutting off the UV modes by means of  an upper extremum  
$\Lambda$ in the integral (\ref{eq: integral}), as done in \cite{eune},
we resort to the more suitable regularization method known as proper time
\cite{Schwinger:1951nm}, and adapt it to the specific 
form of the the integral (\ref{eq: integral}).
Namely, we use the following integral representation of the square 
root appearing above:
\begin{equation}
\int_{0}^{\infty}dsAe^{-s\sqrt{A}}=\sqrt{A}\,,\label{eq: proptrep}
\end{equation}
which is valid for any $A>0$. This assumption is clearly satisfied
in the case of Euclidean version of the Lifshitz-type model,
provided $V'{}'>0$, i.e. we are in the symmetric phase. 
We can take
\begin{equation}
A=p^{4}+\alpha^{-2}V'{}'\,,\label{eq: defA}
\end{equation}
where we rescaled for our convenience
the whole kinetic term by powers of the dimensionful $\alpha$ parameter.
We remind here that we have $[\alpha]=-1$. Moreover, we notice that
the dimension of the regulator in the proper time is $[s]=-2$.
Then, by combining (\ref{eq: proptrep}) with (\ref{eq: defA}) and
plugging this in the integral (\ref{eq: integral}), we get
\begin{equation}
V_{{\rm 1l}}=\frac{\alpha}{4\pi^{2}}\int_{0}^{\infty}ds\!\int_{0}^
{\infty}\!dp\,p^{2}\left(p^{4}+\frac{V'{}'}{\alpha^{2}}\right)
e^{-s\sqrt{p^{4}
+\alpha^{-2}V'{}'}}\,,\label{eq: ptreppot}
\end{equation}
which is the proper time representation of the integral for 
the one-loop potential of our model. We multiplied the
integrand by additional power of the $\alpha$ parameter to comply
with the energy dimension of the potential (in $d=4$ we have 
$[V]=[V_{{\rm 1l}}]=4$). It is important to notice that,
for the success of the proper time regularization program, the
integral over the proper time parameter $s$ must be done at  
last, not to interfere with the integration over spatial momentum.

The momentum  integral in (\ref{eq: ptreppot}) can
be done in an analytic compact form, for a 
general Lifshitz tree-level potential $V=V(\phi)$ 
taken  as a general function of the background scalar field,
through a clever change of integration variable. 
The rough idea is to remove the square root from the exponent 
and the price is that it 
(and its derivatives) will appear in the numerator and also in the 
denominator of the integrand expression. 
We introduce a dimensionless integration variable $y$ (instead
of the dimensionful $p$, with $[p]=1$) defined by the equation
\begin{equation}
\label{eq:y}
\frac{y^{4}V'{}'}{\alpha^{2}}=\frac{V'{}'}{\alpha^{2}}+p^{4}\,,
\end{equation}
where we retain only real positive $p$, so that we can rewrite
Eq. (\ref{eq:y}) as $p=\alpha^{-1/2} \sqrt[4]{y^{4}-1}\;\sqrt[4]{V'{}'}$.
We also assume that the allowed range for the $y$ variable is
$\langle1,+\infty)$ and that  $V'{}'$ and $\alpha$ are naturally positive.
By inserting into the momentum integral the Jacobian ${\cal J}$ of this 
change of integration variables, 
$ {\cal J}={dp}/{dy}=\alpha^{-1/2}  {y^{3}\sqrt[4]{V'{}'}}
{\left(y^{4}-1\right)^{-3/4}}$, we get
\begin{equation}
\int_{0}^{\infty}\!dp\,p^{2}\alpha\left(p^{4}+\frac{V'{}'}{\alpha^{2}}
\right)e^{-s\sqrt{p^{4}+\alpha^{-2}V'{}'}}=\int_{1}^{\infty}\!dy
\frac{y^{7}{V'{}'}^{7/4}\exp\left(-\frac{sy^{2}\sqrt{V'{}'}}
{\alpha}\right)}{\alpha^{5/2}\sqrt[4]{y^{4}-1}}\,.
\end{equation}
By recalling that the integral
\begin{equation}
\label{eq:F}
F(a)=\int_{1}^{\infty}\!dy\frac{y^{7}\exp\left(-ay^{2}\right)}
{\sqrt[4]{y^{4}-1}}
\end{equation}
is expressible in a compact form through the combination of 
Gamma, $\Gamma$,
and Bessel functions, $I$, as
\begin{equation}
F(a)=
\frac{\sqrt{\pi}\Gamma\left(\frac{7}{4}\right)
\left(3I_{-\frac{9}{4}}(a)-2aI_{-\frac{5}{4}}(a)+6I_{\frac{9}{4}}(a)+
2aI_{\frac{13}{4}}(a)\right)}{3\sqrt[4]{2}\, a^{5/4}}\,,\label{eq: intfun}
\end{equation}
we write the final results for the one-loop effective potential in the 
form ($ a={s\sqrt{V'{}'}}/{\alpha}\equiv s\, v>0$)
\begin{eqnarray}
&&V_{{\rm 1l}}=
\frac{1}{4\pi^{2}}\int_{0}^{\infty}\!ds
\frac{{{V'{}'}}^{7/4}}{\alpha^{5/2}}F\left(\frac{s\sqrt{V'{}'}}{\alpha}\right)
\nonumber\\
=\frac{v^{9/4}\alpha\Gamma\left(\frac{3}{4}\right)}{8\sqrt[4]{2}\,\pi^{3/2}}
&&\int_{{0}}^{\infty}\!ds\,s^{-5/4}\left[\frac{3}{2}I_{-\frac{9}{4}}(sv)
+3I_{\frac{9}{4}}(sv)+sv\left(I_{\frac{13}{4}}(sv)-I_{-\frac{5}{4}}(sv)\right)
\right]
\,.
\label{eq: potres}
\end{eqnarray}

Now, the UV divergence is contained in the proper time $s$ integral and
it can be regularized by simply putting a lower cut-off $s_{{\rm UV}}$:
\begin{equation}
\label{eq:regul}
\int_{0}^{\infty}\!ds\to\int_{s_{{\rm UV}}}^{\infty}\!ds\,=
\int_{Bk^{-2}}^{\infty}\!ds\,.
\end{equation}
This cutoff does not act directly on the momentum, but on 
the proper time  $s$, and this allows to bypass  the various 
drawbacks related to the momentum cut-off. Also, in the right hand 
side of (\ref{eq:regul}) we redefine the cutoff $s_{{\rm UV}}$
in terms of a (running) scale $k$, by also including  a free 
constant dimensionless parameter $B$, to be adjusted later, 
namely $s_{{\rm UV}}=Bk^{-2}$. 

The introduction of the cutoff $s_{{\rm UV}}$, makes it possible
to single out the divergent part (in the limit $s_{{\rm UV}}\to 0$)
of the potential, which has the structure of a  sum of inverse 
powers of  $s_{{\rm UV}}$.
Then, the simplest renormalization scheme corresponds 
to the plain subtraction of the divergent terms only, 
via the inclusion of  suitable counterterms. Namely, we perform 
the following subtraction

\begin{eqnarray}
V_{{\rm 1l}}+V_{{\rm ct}}=&&\alpha\!
\int_{0}^{\infty}\!ds\frac{v^{9/4}\Gamma\left(\frac{3}{4}\right)}
{16\sqrt[4]{2}\pi^{3/2}}s^{-5/4}\left[3I_{-\frac{9}{4}}(sv)+6I_{\frac{9}{4}}(sv)
+2 sv\left(I_{\frac{13}{4}}(sv)-I_{-\frac{5}{4}}(sv)\right)\right]
\nonumber\\
&&-\alpha\!\int_{0}^{\infty}\!ds \left(\frac{15}{64\pi^{3/2}}\frac{1}{s^{7/2}}+
\frac{1}{64\pi^{3/2}}\frac{v^{2}}{s^{3/2}}\right) \;.
\label{eq:subtraction}
\end{eqnarray}

The difference of integrals in Eq. (\ref{eq:subtraction}), generated by our 
specific renormalization scheme, can be performed analytically, and the 
output is  (see also \cite{Farias:2011aa})
\begin{equation}
V_{{\rm 1l}}+V_{{\rm ct}}=
\frac{\alpha\Gamma\left(-\frac{1}{4}\right)v^{5/2}}{64\sqrt{2}\pi^{3/2}
\Gamma\left(\frac{9}{4}\right)} \;.
\label{v1lreg}
\end{equation}

\section{Analysis of the effective potential}
\label{sec3}

The result obtained in  Eq. (\ref{v1lreg}) directly yields the full renormalized 
one-loop effective potential that, in the original variables and 
after some manipulation of the Gamma functions, reads

\begin{equation}
V_{{\rm tot}}=V+V_{{\rm 1l}}+V_{{\rm ct}}
=V -\frac{\Gamma\left(\frac{3}{4}\right)^{2}}{10\,\pi^{5/2} \, \alpha^{3/2} }
\, {V'{}'}^{5/4} \;.
\label{vtot}
\end{equation}
Clearly, $V_{{\rm tot}}$ is real only for field values $\phi$ such that 
$V'{}'\geqslant0$, and it precisely reproduces Eq. (29) of \cite{eune}.
\begin{figure}
\begin{centering}
\includegraphics{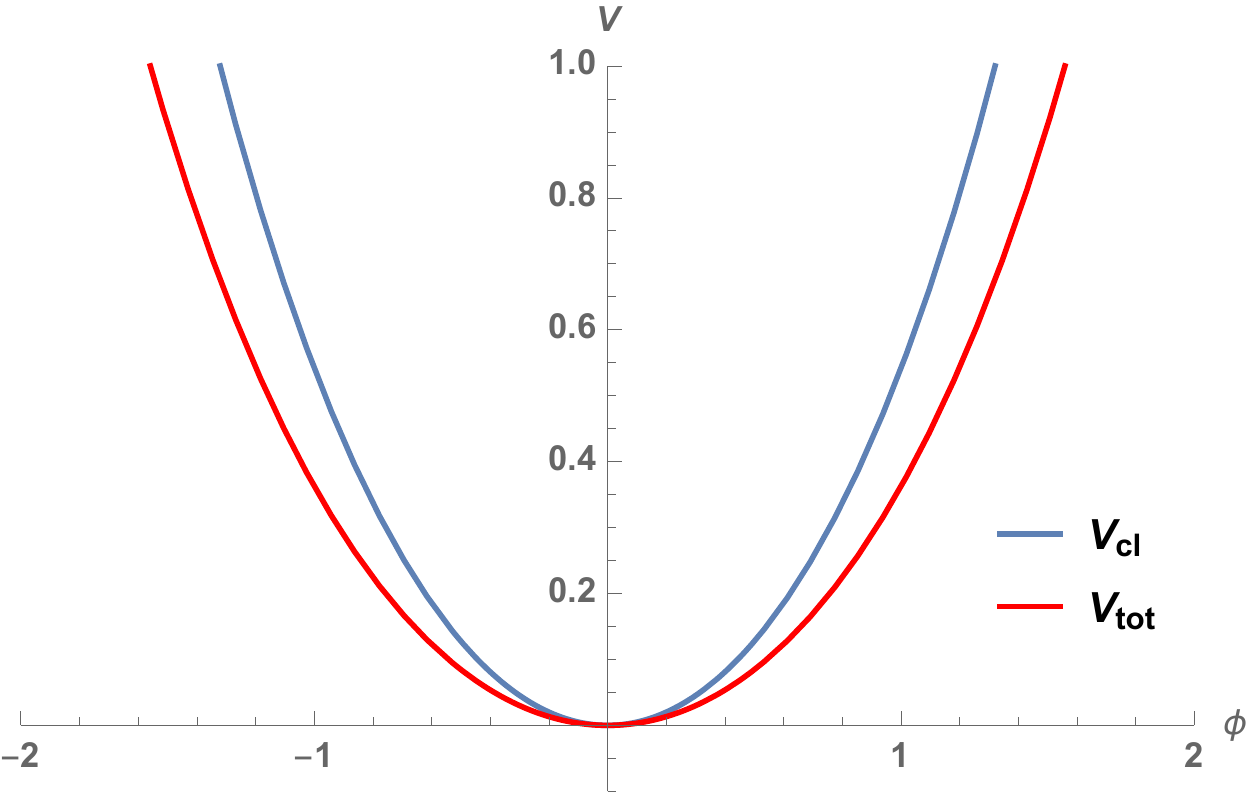}
\par\end{centering}
\caption{\label{fig: plot1} The effective potential  at the classical, 
(upper blue curve) $V$, and one-loop level, (lower red curve) 
$V_{\rm tot}=V_{{\rm 1l}}+V_{{\rm ct}}+V$, plotted
for values of the field $\phi$ around the classical minimum at $\phi=0.$
A field-independent constant is subtracted in $V_{\rm tot}$, in order to get
$V_{{\rm tot}}(\phi=0)=0$.
The values of the parameters are $m^2=\lambda=1$ and $\alpha=0.1$.}
\end{figure}
Now we show in three elucidatory, numerical examples the  comparison
of the tree potential 
\begin{equation}
V=\frac{1}{2}m^2\phi^2+\frac{\lambda}{4!}\phi^4
\label{standardpot}
\end{equation}
with the renormalized one-loop 
potential $V_{\rm tot}$, and we notice that, because of the minus sign in 
front of the loop contribution in Eq. (\ref{vtot}), the quantum 
corrections to $V$ are negative at any value of the field $\phi$.
This is observed in Fig. \ref{fig: plot1}, where $V$ (upper blue curve) and 
$V_{\rm tot}$ (lower red curve) are plotted in a typical configuration 
with no spontaneous symmetry breaking  (SSB), $m^2=\lambda=1$, and $\alpha=0.1$.
(In order to see appreciable differences between the two curves with $m^2$
and $\lambda$ set at 1, it must be $\alpha\ll 1$.)

In addition, for convenience, 
we subtracted to $V_{\rm tot}$ its value at $\phi=0$:
\begin{equation}
V_{{\rm tot}}(\phi=0)= -\frac{m^{5/2}
\Gamma\left(\frac{3}{4}\right)^{2}}{10\,\pi^{5/2} \, \alpha^{3/2} }\;,
\label{v_origin}
\end{equation}
which, for the choice of the parameters used in Fig. \ref{fig: plot1},
is about  $V_{{\rm tot}}(\phi=0)\simeq-0.2715$,
so that for the red curve one observes $V_{{\rm tot}}(\phi=0)=0$.

\begin{figure}
\begin{centering}
\includegraphics{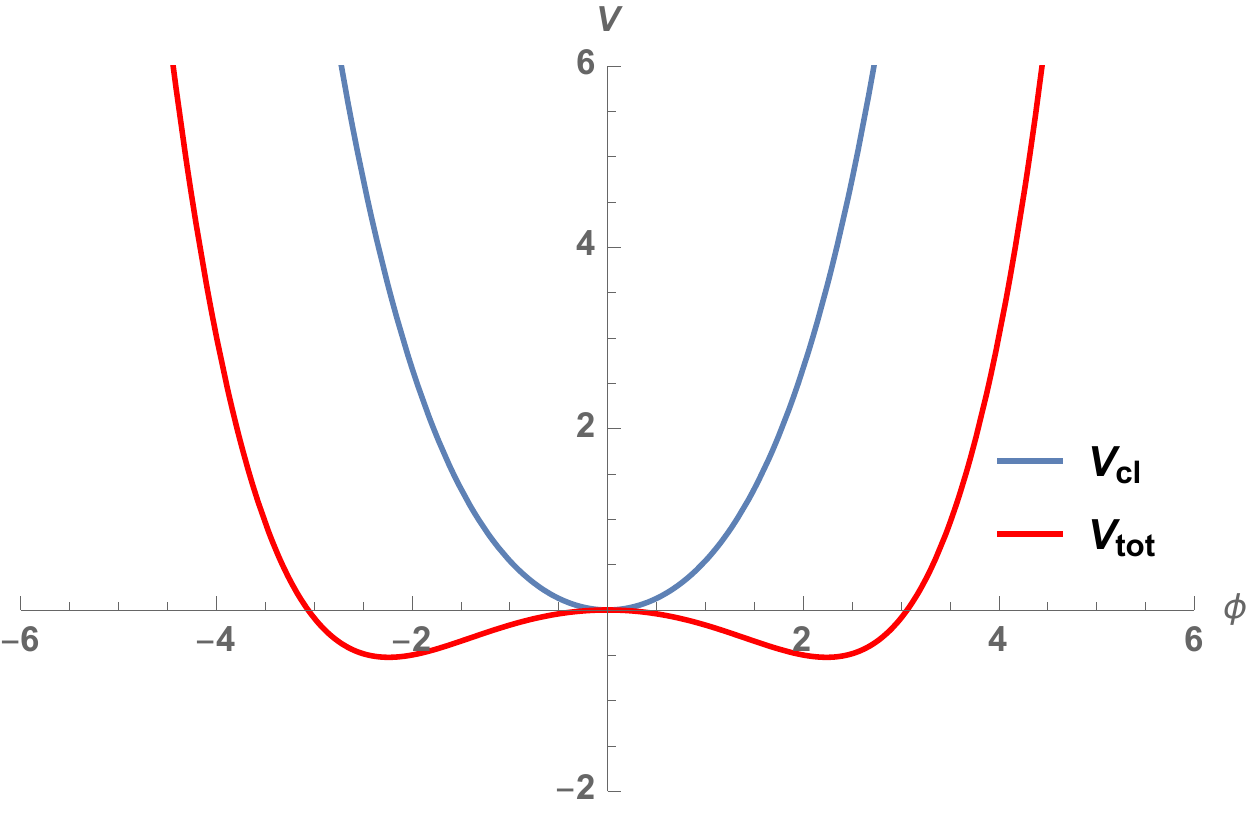}
\par\end{centering}
\caption{\label{fig: plot2}
The effective potential  at the classical, 
(upper blue curve) $V$, and one-loop level, $V_{{\rm tot}}$
(lower red curve).
As in Fig. \ref{fig: plot1}, $m^2=\lambda=1$  and a constant is subtracted from $V_{\rm tot}$, 
so that $V_{{\rm tot}}(\phi=0)=0$.
The value of the remaining parameter is $\alpha=0.04$ and in this case 
$V_{{\rm tot}}$ shows SSB.
}
\end{figure}

For a smaller value of $\alpha$  we observe the generation
of new SSB minima via quantum corrections, for the same tree-level potential. 
This is illustrated in the example in Fig. \ref{fig: plot2} where, 
as in Fig. \ref{fig: plot1} 
$m^2=\lambda=1$, while $\alpha=0.04$.
Again, for the sake of the comparison, 
$V_{{\rm tot}}(\phi=0)\simeq-1.0730$ is subtracted to the one-loop 
effective potential, so that the two curves coincide at $\phi=0$.
Here the full  one-loop curve shows two degenerate absolute minima at non-vanishing
values of the field $\phi$ (and therefore SSB), while $\phi=0$ turns out to be 
a local maximum. Other details of the onset of SSB will be discussed below.

\begin{figure}
\begin{centering}
\includegraphics{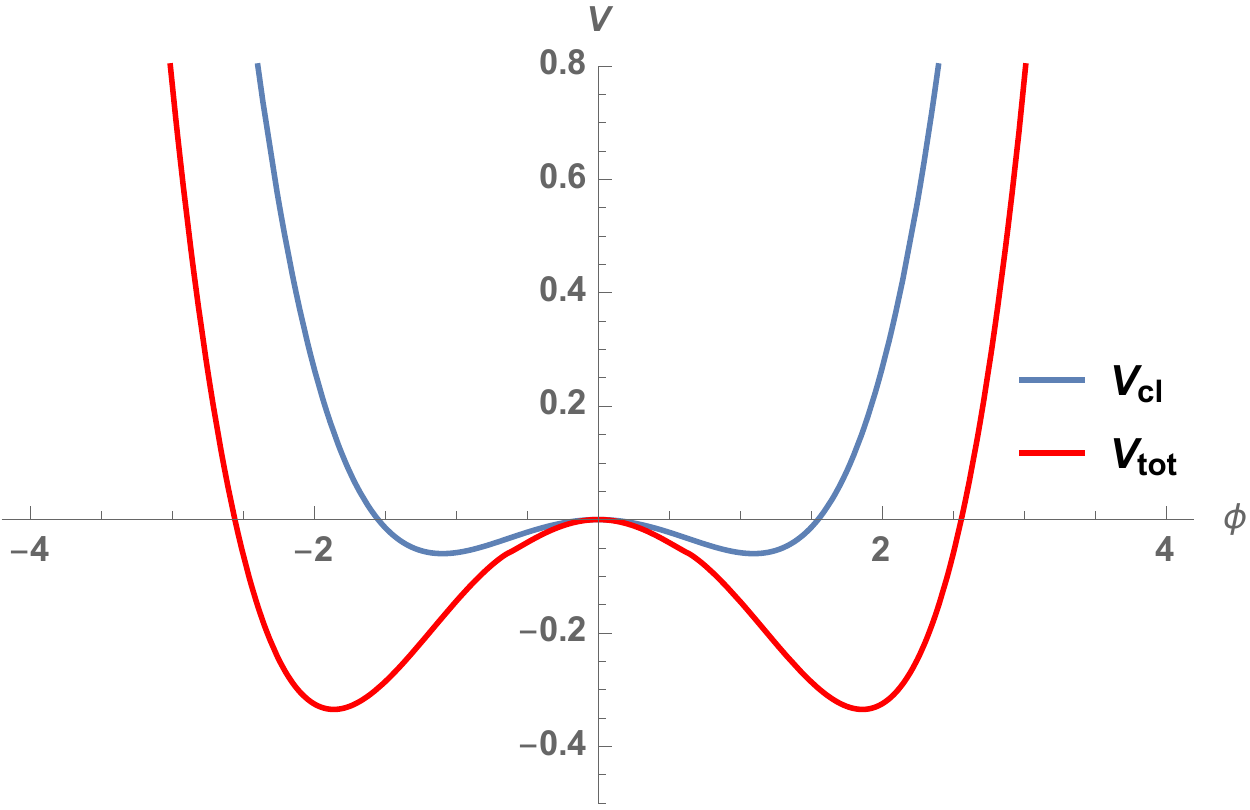}
\par\end{centering}
\caption{\label{fig: plot3}
The real part of effective potential at the classical, 
(upper blue curve) $V$, and one-loop level, $V_{{\rm tot}}$
(lower red curve) with parameters $m^2=-0.2$,
$\lambda=1$ and $\alpha=0.1$. The real part of 
the effective potential at $\phi=0$ is subtracted from $V_{{\rm tot}}$
in order to obtain the lower red curve. In this case both curves show 
SSB.}\end{figure}

Finally, the plot in Fig. \ref{fig: plot3} 
compares the tree-level potential specified by 
$\lambda=1$ and $m^2=-0.2$, and therefore displaying SSB at tree level,
with the real part of the full one-loop potential at $\alpha=0.1$,
which shows even deeper  minima at non-vanishing $\phi$.
It must be remarked that in this case, with negative $m^2$, 
quantum corrections are real at large values of the field, 
but become complex for 
$|\phi|< \sqrt{{-2m^2}/{\lambda}}=\sqrt{ {2}/{5}}\approx0.6325$,
and therefore,  in this region of small field, 
we plot only the real part of the complex total potential $V_{\rm tot}$.
In addition, even in Fig. \ref{fig: plot3} 
the one-loop potential curve is adjusted 
to have a  vanishing effective potential at the origin and this means that 
we subtracted the real value  of Eq. (\ref{v_origin}), which for our choice of the 
parameters is approximately  $0.0257$.

After having shown in the figures the realization of three representative
configurations of the effective potential, we will point out a few properties of
$V_{{\rm tot}}$ that are strictly related to the peculiar form of the action 
(\ref{eq: actionorig}) which generates  the one-loop effective potential 
in Eq. (\ref{vtot}).

The first issue concerns the behaviour of $V_{{\rm tot}}$ at large values of the
field. We notice that, for a generic tree potential 
(i.e. not for the particular potential in Eq. (\ref{standardpot}) )
whose leading term at large $\phi$ is $({g_{q}}/{q!}) \phi^{q}$ (with $g_q >0)$,
the power $q$ must fulfill the condition 
\begin{equation}
q=\frac{5}{4}(q-2) \;,
\label{m10}
\end{equation}
i.e. $q=10$, in order to balance the tree potential and the quantum corrections 
in Eq. (\ref{vtot}), that appear with different sign. If $q>10$, 
$V_{{\rm tot}}$ goes negative at large $\phi$, thus showing instability; 
if $q<10$, $V_{{\rm tot}}$ grows positive at large $\phi$, while if 
$q=10$ one has the critical value  
${g_{10}}_{\rm cr}=4480\pi^{10}\alpha^{6}/
\left({729\,\Gamma\left({3}/{4}\right)^{8}} \right) \approx1.13\cdot10^{5}\alpha^{6}$
above which the effective potential is negative and below which it is positive.
At ${g_{10}}={g_{10}}_{\rm cr}$, the large field behavior of the potential is ruled by 
powers of the field that are smaller than $10$, and therefore it is positive.
Remarkably, as indicated in Eq. (\ref{eq: potentialtree}), $\phi^{10}$ 
corresponds to the marginally scaling operator according to the Lifshitz scaling.

Now we investigate on the condition of masslessness of the effective potential at
$\phi=0$, which is specified by the condition (the subscript $0$ means that 
it is evaluated at $\phi=0$)
\begin{equation}
{m_0}_{{\rm eff}}^{2}=\left.\frac{d^{2}V_{{\rm tot}}}{d\phi^{2}}\right|_{\phi=0}=0\;.
\label{masslessness}
\end{equation}
This issue is strictly related to the presence of a non-trivial minimum at 
$\phi\neq 0$, as it is clear that a~negative curvature (mass square parameter) at $\phi=0$ for a
well-behaved effective potential that diverges positively at large $\phi$,
indicates the presence of minima at $\phi\neq 0$. 

We start by considering 
a simple monomial tree potential $V=(g_{q}/{q!}) \phi^{q}$ with $q>2$, so that 
we assume for the moment a zero tree mass in our problem. By looking at the 
structure of $V_{{\rm tot}}$ in Eq. (\ref{vtot}), we immediately realize that 
we get ${m_0}_{{\rm eff}}^{2}\neq 0$ only if $(5/4)(q-2)=2$, that selects the 
particular value $q=18/5$. Instead, if we take a monomial potential with $q>18/5$,
we find ${m_0}_{{\rm eff}}^{2}=0$ and, if $q<18/5$, the curvature at the origin 
diverges. If we limit ourselves to integer $q$, then the first value that does not 
generate singularities at the origin is $q=4$. In general, integer $q$, with
$4\leqslant q \leqslant 10$ generate regular effective potentials with 
${m_0}_{{\rm eff}}^{2}=0$.

We go one step further and search for non-trivial minima of the effective
potential in the case of monomial tree potential $V=(g_{q}/{q!}) \phi^{q}$
and, with the help of Eq. (\ref{vtot}), we find in this case
\begin{equation}
V_{{\rm tot}}=\frac{g_{q}}{q!}\phi^{q}-\gamma\left(\frac{g_{q}}{(q-2)!}\phi^{q-2}
\right)^{5/4},
\label{vmonomtot}
\end{equation}
where, for the sake of simplicity, we defined 
\begin{equation}
\gamma=\frac{\Gamma\left(\frac{3}{4}\right)^{2} }{10\pi^{5/2}\alpha^{3/2}}  
\label{gamma}
\end{equation}
and the extremum condition $V_{{\rm tot}}'=0$ admits, besides $\phi=0$, 
the non-zero solution
\begin{equation}
\phi_{{\rm min}}=\left(\frac{5(q-1)!\gamma
g_{q}^{1/4}}{4((q-2)!)^{1/4}(q-3)!}\right)^{\frac{4}{10-q}}\;,
\label{phimin}
\end{equation}
provided that $4\leqslant q <10$.
This solution, as discussed above, must be a minimum, in opposition to the maximum 
at the origin.
The negative value of the effective potential at $\phi_{{\rm min}}$
can be straightforwardly computed and, in particular, we display the results 
for $q=4$
\begin{equation}
\phi_{{\rm min}}=\left(\frac{15\gamma g_{4}^{1/4}}{2^{5/4}}\right)^{\frac{2}{3}}
=
\left(\frac{3\Gamma\left(\frac{3}{4}\right)^{2} g_{4}^{1/4}}{2^{9/4}\pi^{5/2}
\alpha^{3/2}}\right)^{\frac{2}{3}}
\label{phimin4}
\end{equation}
and
\begin{equation}
V_{{\rm tot}}(\phi_{{\rm min}})
=- \; \frac{45 \;  5^{2/3}3^{2/3}}{32 \; 2^{1/3}} 
g_{4}^{\frac{5}{3}}\gamma^{\frac{8}{3}} \;.
\label{Vmin4}
\end{equation}

Therefore the structure of Eq. (\ref{vtot}) with a monomial tree potential 
with $4\leqslant q < 10$ and  vanishing tree mass, $m^2=0$, 
yields a double well effective potential with zero curvature at the origin.
One should be aware that this picture is somehow different
from the one of the Coleman-Weinberg one-loop scalar potential, 
where the only dimensionful scale is generated  by the 
radiative corrections, while in our case (even when $q=4$ and
the coupling $g_4$ is dimensionless) we start with at least 
one dimensionful input parameter, namely $\alpha$.

At this point we go back to the study of the curvature of the effective potential
at $\phi=0$, but now we allow for a finite tree mass term, i.e. we take the tree
potential as 
\begin{equation}
V=\frac{1}{2}m^{2}\phi^{2}+\frac{g_{q}}{q!}\,\phi^{q}
\label{Vpoly}
\end{equation}
with $m^2>0$, and expand the second derivative of the effective potential 
in Eq. (\ref{vtot}) as
\begin{equation}
{m_0}_{{\rm eff}}^{2}
=\left.\frac{d^{2}V_{{\rm tot}}}{d\phi^{2}}\right|_{\phi=0}=
m^2  - \gamma \left (\frac{5}{16 \,m^{3/2}} \lim_{\phi\to 0}{V'''}^{2}+
\frac{5\,m^{1/2}}{4}\lim_{\phi\to 0}V'''' \right ) \;.
\label{eqmeff}
\end{equation}

Then,  if $2<q<3$ or $3<q<4$, a singularity 
shows up in the right hand side of Eq. (\ref{eqmeff}) and therefore, 
in order  to have a regular effective potential, we must take 
either $q=3$ or $q\geqslant 4$. In addition, when $q=3$ and $q=4$, 
there is a finite quantum correction to  $m^2$
while, if $q>4$, the correction  vanishes and ${m_0}_{{\rm eff}}^{2}=m^2$.
For $q=3$  we get
\begin{equation}
{m_0}_{{\rm eff}}^{2}
=m^{2}-
\frac{\Gamma\left(\frac{3}{4}\right)^{2}}{32\pi^{5/2}\alpha^{3/2}} 
\frac{g_{3}^{2}}{m^{3/2}}
\label{mtre}
\end{equation}
but, as  we are mainly interested in even potentials, we do not further analyze
the case $q=3$ and we look instead at $q=4$, where we get, according to the definition (\ref{gamma}),
\begin{equation}
{m_0}_{{\rm eff}}^{2}
=m^{2}-
\frac{\Gamma\left(\frac{3}{4}\right)^{2}}{8\pi^{5/2}\alpha^{3/2}}m^{1/2} g_4 =
m^{2}-
\frac{  5\,\gamma \, m^{1/2}\, g_4}{4}   
\;.
\label{mquattro}
\end{equation}

Eq. (\ref{mquattro}) shows that the sign of ${m_0}_{{\rm eff}}^{2}$ is determined 
by the relative size of the parameters $m^2$, $\alpha$ and $g_4$. 
In particular, by defining 
\begin{equation}
C (m^2)=  \frac{5\,\gamma \,  g_4}{4 \,  m^{3/2} }  \;,
\label{mquattrozero}
\end{equation}
we find  that ${m_0}_{{\rm eff}}^{2}=0$
occurs both at $m^2=0$ and at $m^2=\overline m^2$, where $\overline m^2$ is 
such that $C(\overline m^2)=1$.

For large positive $m^2$, the tree mass is 
dominant in Eq. (\ref{mquattro}) and therefore 
${m_0}_{{\rm eff}}^{2}>0$ and it corresponds to $0<C<1$. Conversely,
for small positive $m^2$, i.e. $C>1$,
the negative quantum correction is dominant in Eq. (\ref{mquattro}) 
and we have ${m_0}_{{\rm eff}}^{2}<0$.

Therefore,
for large $m^2$ ( $0<C<1$) the curvature at the origin, as expected, is positive
(a particular example of this  configuration is given in Fig. \ref{fig: plot1})
but, more interestingly, we find a full interval of $m^2$, from $m^2=0$ ($C\to \infty$) 
to the critical value $\overline m^2$ associated to $C=1$, 
that yields negative curvatures at $\phi=0$ (with vanishing 
curvature at the two endpoints of the interval) and consequently implies the 
onset of a new couple of  minima at $\phi\neq 0$.

In fact, the case with $m^2=0$ has been discussed 
before in the context of monomial potentials and the corresponding 
minima associated to SSB have been determined. 
When $m^2$ grows from zero, the curvature at the origin diminishes from
zero and becomes negative, and the corresponding  SSB minima can be 
determined numerically (Fig. \ref{fig: plot2} is an example of this kind).
The location of these minima approaches zero when $m^2\to \overline m^2$
and they disappear for  $m^2> \overline m^2$, 
which marks the transition to the symmetric phase with the only 
minimum of the effective potential located at $\phi=0$.

In summary, the picture observed for this Lifshitz-type action 
is rather different from that of the simple scalar effective potential,
as in the former case there is a finite range of values of $m^2>0$ 
that produces negative curvature at $\phi=0$ and SSB with non-trivial 
minima, while in the latter case the negative curvature is obtained
only for $m^2 < 0$. In addition, in the former case  there are two
different values, namely $m^2=0$ and $m^2=\overline m^2$, associated to
zero curvature ${m_0}_{{\rm eff}}^{2}=0$. As already noticed,
these differences are essentially due to the presence of more than one 
dimensionful parameter in our problem that give origin to a richer structure.

Finally, the case with $m^2<0$, that corresponds to a SSB 
tree level potential,
presents the problem of  complex quantum corrections (at least 
for small values of the field $\phi$), due to the term 
${V''}^{5/4}$ in Eq. (\ref{vtot}). 
In this case, as expected, a couple of minima 
at $\phi\neq 0 $ is always present and $V_{{\rm tot}}(\phi_{{\rm min}})$ 
has zero imaginary part. 
In our analysis, we focused only on the real part of 
$V_{{\rm tot}}$ and a specific example of this kind is displayed in 
Fig. \ref{fig: plot3}.

\section{Flow of the effective potential and beta-functions}
\label{sec4}

After analysing the renormalized one-loop effective potential, 
we now turn to the 
issue of determining  the flow equation for the scale-dependent potential.
To this purpose, we recall that the standard procedure is obtained 
by applying
$k_{\rm IR}\, d /dk_{\rm IR}$ to the effective potential regulated by an 
infrared scale $k_{\rm IR}$ and, in our case, it is easy to realize that this is
equivalent to applying $- k\, d /dk$ to Eq. (\ref{eq: potres}),
properly regularized in accordance to 
Eq. (\ref{eq:regul}), and the factor  $(-1)$  is introduced 
to compensate the exchange of the UV and IR cutoff. Therefore, we get
\begin{eqnarray}
k\frac{d}{dk}V_{{\rm 1l}}(k)=&& -
\frac{{V''}^{9/8}}{4\sqrt[4]{2}\pi^{3/2}\alpha^{9/4}}\frac{k^{1/2}}
{B^{1/4}}\left[2\alpha\Gamma\left(\frac{7}{4}\right)
\left(I_{-\frac{9}{4}}\left(\frac{B\sqrt{V'{}'}}{\alpha
k^{2}}\right)+2I_{\frac{9}{4}}\left(\frac{B\sqrt{V'{}'}}{\alpha
k^{2}}\right)\right)\right.
\nonumber\\
&&\left.+\frac{B}{k^{2}}\Gamma\left(\frac{3}{4}\right)\sqrt{V'{}'}
\left(I_{\frac{13}{4}}\left(\frac{B\sqrt{V'{}'}}{\alpha 
k^{2}}\right)-I_{-\frac{5}{4}}\left(\frac{B\sqrt{V'{}'}}{\alpha 
k^{2}}\right)\right)\right].
\label{eq: flow}
\end{eqnarray}

From the above expression one could get the naive expectation that the
flow of the potential is always proportional to the square root 
$\sqrt{k}$. However, this is not true, because, for
example, in the large $k$ regime we also have to expand in series
the dependence on $k$ in the argument $x$ of the Bessel
functions. It is known that when $x\to 0$, one has
\begin{equation}
\label{eq:bessel}
I_{b}(x)\underset{x\to0}{\sim}\frac{1}{\Gamma(1+b)}
\left(\frac{x}{2}\right)^{b}+O\left(x^{b+2}\right)
\end{equation}
and Eq. (\ref{eq:bessel}) allows to establish the UV regime
 ($x={B\sqrt{V'{}'}}/({\alpha k^{2}})\to0$)
of the RG flow of the potential, where it is assumed $k\gg\alpha^{-1}$
and only the two leading powers of $k$ are 
retained:

\begin{equation}
\label{eq:UVflow}
k\frac{d}{dk}V_{{\rm 1l}}(k)\underset{{\rm UV}}{=} -
\frac{15}{32\pi^{3/2}}\frac{\alpha}{B^{5/2}}k^{5} -
\frac{V''k}{32\pi^{3/2}\alpha\sqrt{B}}+O\left(k^{-3/2}\right)\,.
\end{equation}

The flow in the IR regime can be obtained similarly:

\begin{equation}
\label{eq:UVflow2}
k\frac{d}{dk}V_{{\rm 1l}}(k)
\underset{{\rm IR}}{=} -
\frac{\Gamma(\frac{3}{4})B^{1/4} {V''}^{11/8}\exp \left 
(-\frac{B\sqrt{V'{}'}}{\alpha k^2} \right )}
{4\pi^2\sqrt[4]{2} \, \alpha^{7/4}\sqrt{k} }\,.
\end{equation}
This flow is a non-analytic function of $k$ and goes to zero like
$\exp(-1/k^{2})$, so that the running is exponentially 
dumped at low momenta.

Now, we derive the $\beta$-functions of some relevant couplings 
from Eq. (\ref{eq: flow}). The couplings are defined by 
the parametrization of potential in  (\ref{eq: potentialtree}) 
and this time, for illustrative purposes, we focus
on the first three parameters,
namely $m^{2}(k)$, $\lambda(k)\equiv \lambda_1(k)$ and
$g(k)\equiv \lambda_2(k)$.
Their $\beta$-functions are obtained by projecting the
flow of the potential onto the specific subspace of each coupling
through successive differentiation of the right hand side of 
Eq. (\ref{eq: flow})  with respect to the field.

Then, for the square mass parameter we find 
($w\equiv (B m)/(\alpha k^{2})$ )

\begin{eqnarray}
\label{mbeta}
&&\beta_{m^{2}}=
\frac{\lambda B^{\frac{7}{4}}} 
{32 \,\pi^{\frac{3}{2}}\,\alpha^{\frac{13}{4}} \,k^{\frac{7}{2}}}
\, \left(\frac {m}{2} \right)^{\frac{1}{4}}\,
\left[2\Gamma\left(\frac{3}{4}\right)\left(
\frac{27\alpha^{2}k^{4}}{B^2}+
2m^{2}\right)I_{-\frac{9}{4}}\left( w \right)\right.
\nonumber\\&&
\left.+m\Gamma\left(-\frac{1}{4}\right)\left(\frac{3\alpha k^{2}}{B}
\left(I_{\frac{5}{4}}\left(w\right)-
I_{-\frac{13}{4}}\left(w\right)\right)
+m\, I_{\frac{9}{4}}\left(w\right)\right)\right]
\underset{{\rm UV}}{=}-\frac{k\lambda}{32\,\pi^{\frac{3}{2}}\,
\alpha\, \sqrt{B}}\, ,
\end{eqnarray}
where the right hand side shows the result obtained 
in the UV regime, i.e. with $k\gg\alpha^{-1},m$. In fact,
in this limit many irrelevant details associated to the 
particular nature of the IR regulator become negligible.
We observe that $\beta_{m^{2}}$ is linearly proportional to the
quartic coupling $\lambda$. 

We also notice that $\beta_{m^{2}}$
corresponds to a dimensionful  parameter 
($[m^{2}]=2)$, while the $\beta$-function of the associated
dimensionless coupling 
$\tilde{m}^{2}={m}^{2}B/k^{2}$ is:
$\beta_{\tilde{m}^{2}}=-2\tilde{m}^{2}+B\,k^{-2}\beta_{m^{2}}$,
where the first term is generated from the dimension of the 
original coupling $m^{2}$.

Similarly, one derives $\beta_{\lambda}$ of the dimensionless
quartic coupling $\lambda$, displayed below 
only in its simplified form in the UV regime. 
It turns out 
to be proportional to the sextic coupling $g\equiv\lambda_2$:
\begin{equation}
\beta_{\lambda}\underset{{\rm UV}}{=}
-\frac{g k}
{32\,\pi^{\frac{3}{2}}\,\alpha\, \sqrt{B}}\, =
-\frac{\tilde{g}} {32\,\pi^{\frac{3}{2}}\,\tilde{\alpha} }
\label{eq: betalamUV}
\end{equation}
and, in the right hand side of  (\ref{eq: betalamUV}), 
$\beta_{\lambda}$  is expressed in terms of dimensionless variables
$\tilde{g}=g k^{2}/B$ and $\tilde{\alpha}=\alpha k/\sqrt{B}$.
The same procedure, carried out for the other relevant 
couplings, shows again
that $\beta_{\lambda_2}\propto \lambda_3$  and 
$\beta_{\lambda_3}\propto \lambda_4$.

Note that we retained the factor $B$, introduced in Eq. 
(\ref{eq:regul}), into the definition of running scale and 
in the consequent definition of the above 
dimensionless variables, as $B$ is an indispensable 
element in the rescaling procedure. 
Then, only within this scheme and in the UV limit,
one recovers universal expressions
(i.e. independent of the details of the IR regulator, 
such as $B$)
for the one-loop 
$\beta$-functions of the dimensionless couplings.

\section{Conclusions}
\label{sec5}
In conclusion, we developed a new regularization scheme, suitable for 
studying the physics around anisotropic Lifshitz points, by 
a standard treatment of the time coordinate and
by adapting the proper time regulator to the 
three-dimensional subset of space coordinates, by means of the 
integral representation of the square root  in 
(\ref{eq: proptrep}). Then,  the UV divergences are regulated 
by a sharp cutoff on the proper time variable $s$. 
With the help of  this scheme, we computed the one-loop effective 
potential, by determining the correct counterterms to get finite 
quantum corrections, which turn out to decrease the value of the 
tree potential at each value of the field $\phi$, as shown 
in the three figures. 
We also pointed out that the presence in our problem  of two or more
dimensionful scales gives origin to a rich structure of the phase diagram,
which allows for a full interval of positive values of the tree level square 
mass that are  associated to a negative curvature at the origin,
with consequent onset of SSB at the one-loop level.

Then, from the dependence of the potential and its couplings
on the cutoff scale, we derived a flow equation for the 
effective potential and the $\beta$-functions for  the couplings
$m^{2}$ and $\lambda$,
and we found that
$\beta_{\lambda_n}$ is proportional to  the subsequent coupling
$\lambda_{n+1}$, in agreement with the  findings of \cite{eune}.

The $\beta$-functions, 
are rather different from those
calculated for the theory in proximity of the Gaussian fixed point, because
they are a consequence of the nature of the ultraviolet divergences associated
to a Lifshitz point. In fact, an inspection of the
diagrams within the Lifshitz scaling indicates only one divergent 
one-loop diagram for each  $(2n+2)$-point Green's function,
namely the tadpole generated by the coupling $\lambda_{n+1}$.

Moreover, a comparison with the counterterms determined in \cite{eune},
allows us to select a particular value of the parameter $B$,
i.e. the proportionality constant between the proper time
cutoff $s_{\rm UV}$ and the inverse square running scale $k^{-2}$.
Actually, by choosing 
$\sqrt{B}={\sqrt{\pi}}/{4}$
in (\ref{mbeta}) and (\ref{eq: betalamUV}), we get,
${\beta}_{m^{2}}=-k\lambda/({8\pi^{2}\alpha})$,
${\beta}_{\lambda}=-{k}g/({8\pi^{2}\alpha})$,
that reproduce the results  obtained 
with a sharp 3-momentum UV cutoff in \cite{eune}.

\vspace{1cm}
{\it Acknowledgements }: AB thanks Ugo Moschella for important comments. MP was supported by a KIAS Individual Grant (PG062001) at Korea Institute for Advanced Study and by Basic Science Research Program through the National Research Foundation of Korea funded by the Ministry of Education (NRF-2016R1D1A1B03933399). 
The work of LR was partially supported by the ACRI-INFN Research Award within Young Investigator Training Program 2018 in the project ``Functional and Renormalization-Group Methods in Quantum and Statistical Physics''. MP and LR would like to thank INAF-Catania Astrophysical Observatory for kind hospitality.

\bibliographystyle{jhepbib}
\bibliography{proper}
\end{document}